\documentclass[12pt]{article}
\usepackage[english,german,french,polish]{babel}
\usepackage[T1]{fontenc}
\usepackage{amsfonts}

\begin{document}

\selectlanguage{english}

\baselineskip 0.75cm
\topmargin -0.6in
\oddsidemargin -0.1in

\let\ni=\noindent

\renewcommand{\thefootnote}{\fnsymbol{footnote}}

\newcommand{\SM}{Standard Model }

\pagestyle {plain}

\setcounter{page}{1}

\pagestyle{empty}

~~~

\begin{flushright}
IFT--05/25
\end{flushright}

\vspace{1.0cm}

{\large\centerline{\bf A very simple empirical neutrino mass formula{\footnote{Work supported in part by the Polish State Committee for Scientific Research (KBN), grant 1 P03B 099 29.}}}}

\vspace{0.4cm}

{\centerline {\sc Wojciech Kr\'{o}likowski}}

\vspace{0.3cm}

{\centerline {\it Institute of Theoretical Physics, Warsaw University}}

{\centerline {\it Ho\.{z}a 69,~~PL--00--681 Warszawa, ~Poland}}

\vspace{0.6cm}

{\centerline{\bf Abstract}}

\vspace{0.2cm}

A very simple empirical neutrino mass formula is found, implying the mass proportion
$m_1 : m_2 : m_3 = 1 : 4 : 24(1-\beta^2)$. For the value $\beta = (5-1)(5-3)/5^2 = 8/25$ the 
formula predicts precisely $\Delta m^2_{21} \sim 8.0\times 10^{-5}\;{\rm eV}^2$ in consistency 
with the experiment, when the input of experimental estimate $\Delta m^2_{32} \sim 
2.4\times 10^{-3}\;{\rm eV}^2$ is applied. Plausible interpretation of numerical elements  
building up this formula is briefly described.

\vspace{0.2cm}

\ni PACS numbers: 12.15.Ff , 14.60.Pq , 12.15.Hh .

\vspace{0.6cm}

\ni October 2005 

\vfill\eject

~~~
\pagestyle {plain}

\setcounter{page}{1}

\vspace{0.2cm}

The role of empirical spectral formulae in the history of atomic and nuclear physics cannot be overestimated. The Balmer formula for hydrogen spectrum is perhaps the most famous example. At the advent of quantum physics, it has inspired the semiclassical Bohr model of hydrogen atom and, in a further consequence, the quantum mechanics of Schr\"{o}dinger and Heisenberg. In particle physics, attempts to construct empirical spectral formulae for leptons and quarks are not very popular ({\it cf.} Refs. [1,2] for some recent publications). The main reason may be the apprehension about the possibility of not controlled speculations connected with the fact that, on the level of fundamental particles, the spectral formulae become actually mass formulae, while the very concept of mass has not been advanced essentially from Newton's time, in contrast to the enormous progress in particle dynamics operating with masses having still the status of free parameters (in Higgs mechanism in the \SM and its extensions, the corresponding Yukawa coupling constants are free parameters). But, paradoxically, it seems that in the present situation, where the mass concept waits to be developed, the search for empirical mass formulae for leptons and quarks ought to be natural.

In the present note, we report on an efficient empirical mass formula for three mass neutrinos $\nu_1, \nu_2 , \nu_3$ related to three active neutrinos $\nu_e \,,\, \nu_\mu \,,\, \nu_\tau $ through the unitary transformation $\nu_\alpha = \sum_i U_{\alpha i} \nu_i \;(\alpha = e, \mu, \tau$ and $ i =1,2,3)$. Making use of a numerical experience, we conjecture the following neutrino mass formula:

\begin{equation}
m_i  =  \mu\, \rho_i (1 - \beta^2 \delta_{i 3}) \;\;(i = 1,2,3)
\end{equation}

\ni with

\begin{equation}
\beta^2 \delta_{i3} = \left[\frac{(N_i - 1)(N_i - 3)}{N^2_i} \right]^2 = \left(\frac{8}{25}\right)^{\!\!2}\! \delta_{i 3} = 0.1024 \,\delta_{i 3} \,,
\end{equation}

\ni where $ N_1 = 1, N_2 = 3, N_3 = 5 $ are three odd integers and $ \rho_1 = 1/29 , \rho_2 = 4/29 ,
\rho_3 = 24/29 $ --- three normalized fractions ($\sum_i \rho_i = 1$) that may be called 
generation-weighting factors. Here, $\mu > 0$ is a constant to be determined from, say, the experimental estimate of $\Delta m^2_{32} \sim 2.4\times 10^{-3} \;{\rm eV}^2 $ [3]. In the formula (1), the masses $m_1, m_2 , m_3 $ depend linearly on the generation-weighting factors $ \rho_1, \rho_2 ,\rho_3 $, respectively. The mass $m_3$ of the neutrino $\nu_3$ of the third generation involves also an additive correction proportional to the value $\beta^2 =$ $\left[(N_3 -1)(N_3-3]/N_3^2\right]^2 = (8/25)^2 = 0.1024 $ that is pretty small {\it versus} 1.

The neutrino mass formula (1) shows that  

\begin{equation}
m_1 : m_2 : m_3 = 1 : 4 : 24(1-\beta^2)
\end{equation}

\ni and

\begin{equation}
\mu = 29 m_1 = \frac{29}{4} m_2 = \frac{29}{24(1-\beta^2)} m_3 = \frac{29}{5+24(1-\beta^2)} (m_1 + m_2 + m_3) \,.
\end{equation}

\ni From Eq. (3) it follows that 

\begin{equation}
\Delta m^2_{32} = \mu^2 \frac{16}{841}\left[36(1 - \beta^2)^2 - 1 \right] = 0.5328\, \mu^2
\end{equation}

\ni and

\begin{equation}
\Delta m^2_{21} = \mu^2 \frac{15}{841} = 0.01784\, \mu^2\,.
\end{equation}

\ni Thus,

\begin{equation}
\frac{\Delta m^2_{32}}{\Delta m^2_{21}} = 29.87\,.
\end{equation}

\ni The input $\Delta m^2_{32} \sim 2.4\times 10^{-3} \;{\rm eV}^2 $ [3] used in Eq. (7) gives the {\it prediction}

\begin{equation}
\Delta m^2_{21} \sim 8.0\times 10^{-5} \;{\rm eV}^2,
\end{equation}

\ni precisely consistent with the experimental estimate of $\Delta m^2_{21}$ [3]. The same input applied in Eq. (5) leads to the fitting

\begin{equation}
\mu  \sim 6.7\times 10^{-2} \;{\rm eV},
\end{equation}

\ni and then {\it predicts} from Eqs. (4) that

\begin{equation}
m_1 \sim 2.3\times  10^{-3}\; {\rm eV} \;,\;m_2 \sim 9.3\times  10^{-3}\; {\rm eV} \;,\;m_3 \sim 5.0\times  10^{-2}\; {\rm eV} \,.
\end{equation}

Concluding, we can see that the {\it very simple} mass proportion (3) implied by our neutrino mass formula (1) {\it predicts} precisely the correct ratio of the mass-squared scales appearing in the so different phenomena as the atmospheric and solar neutrino oscillations. For this correct prediction the choice of $\beta^2 \simeq 0.10 $ is necessary. Such a choice is provided all right in the case of our conjecture $\beta^2\, \delta_{i 3} = [(N_i -1)(N_i-3]/N_i^2]^2 = (8/25)^2\, \delta_{i 3} = 0.1024\, \delta_{i 3} $.

It is exciting enough to observe that an efficient empirical mass formula for charged leptons found out previously [2] is built up of {\it the same} numerical elements $N_i$ and $\rho_i \;(i=1,2,3)$ as the neutrino mass formula (1).

Although the formula (1) has essentially an empirical character, its attractive simplicity can be connected with a speculative background based on a K\"{a}hler-like extension of the Dirac equation ({\it i.e.} on an extension of Dirac's square-root procedure) which the interested reader may find in Ref. [2]. In particular, the  numerical elements $N_i$ and $\rho_i \;(i=1,2,3)$ present in formula (1) can be interpreted there. In fact, $N_i - 1 = 0,2,4$ is there the number of {\it additional} bispinor indices appearing in the extended Dirac equation and obeying Fermi statistics (along with Pauli principle) that enforces their antisymmetrization and so, restricts to zero the related additional spin.  This antisymmetrization is also the reason, why there are precisely {\it three} \SM fermion generations {\it i.e.}, $N_i - 1 = 0,2,4$ (since any of $N_i -1$ additional bispinor indices can assume four values, what then implies that $N_i -1 \leq 4$). Thus, an analogue of Pauli principle works here {\it intrinsically} for $N_i -1$ additional bispinor indices treated as physical objects. Besides, there is one bispinor index, distinguishable from those additional, that must be associated with the set of $SU(3)\times SU(2)\times U(1)$ labels numerating 15+1 states in each \SM generation. So, there are altogether $N_i = 1,3,5 $ bispinor indices treated as physical objects that can be called {\it algebraic spin-1/2 partons} of leptons and quarks. The generation-weighting factors $\rho_i = 1/29, 4/29, 24/29 $ multiplied by 29 tell us {\it how many times} the lepton or quark wave-functions of three generations {\it are realized} (up to the factor $\pm1$) in the extended Dirac equation.

The above picture of leptons and quarks is {\it intrinsic} in the sense that the introduced spin-1/2 partons are entirely algebraic objects, not related to any individual positions within leptons and quarks. Such a picture of algebraic compositness abstracts from the spatial notion of familiar compositness and remains in an ideological analogy to the Dirac intrinsic picture of spin 1/2 that abstracts from the spatial notion of familiar angular orbital momentum. In our opinion, this picture of leptons and quarks is especially attractive in its {\it pure} form characterized above. It may happen, however, that in reality the bispinor indices appearing in this picture show up only the summit of a hidden iceberg consisting of more familiar {\it spatial spin-1/2 partons} bound (usually) in orbital S states.

Of course, in spite of some hints, the conjectured neutrino mass formula (1) cannot be justified on the ground of the present particle theory that, strictly speaking, does not include any predictive mass theory. The role of this empirical mass formula is just to help in finding the adequate mass theory, at least in the neutrino sector. Similarly, the empirical Balmer formula for hydrogen spectrum could not be justified on the ground of prequantal theory but, nevertheless, it inspired the discovery of Bohr semiclassical quantum conditions.

\vfill\eject

~~~~

\vspace{0.8cm}

{\centerline{\bf References}}

\vspace{0.5cm}

{\everypar={\hangindent=0.7truecm}
\parindent=0pt\frenchspacing

{\everypar={\hangindent=0.7truecm}
\parindent=0pt\frenchspacing

[1]~For a recent discussion of the Koide equation for charged leptons {\it cf. } Y.~Koide, {\tt hep--ph/0506247}; {\tt hep--ph/0509214}; and references therein.

\vspace{0.2cm}

[2]~For a recent presentation of an efficient charged-lepton mass formula {\it cf.} W. Kr\'{o}likowski, {\it Acta Phys. Pol.} {\bf B 33}, 2559 (2002) [{\tt hep--ph/0203107}]; {\tt hep--ph/0504256}; 
{\tt hep--ph/0509043}; and references therein.

\vspace{0.2cm}

[3]~{\it Cf. e.g.} G.L. Fogli, E. Lisi, A. Marrone and A. Palazzo, {\tt hep--ph/0506083}.

\vfill\eject

\end{document}